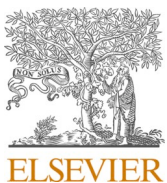
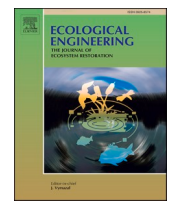

# Cost-benefit of green infrastructures for water management: A sustainability assessment of full-scale constructed wetlands in Northern and Southern Italy

Laura García-Herrero [a,*], Stevo Lavrnić [a], Valentina Guerrieri [a], Attilio Toscano [a], Mirco Milani [b], Giuseppe Luigi Cirelli [b], Matteo Vittuari [a]

[a] *Department of Agricultural and Food Sciences, Alma Mater Studiorum-University of Bologna, Viale Giuseppe Fanin 50, 40127 Bologna, Italy*
[b] *Department of Agriculture, Food and Environment (Di3A), University of Catania, Via Santa Sofia No. 98-100, 95123 Catania, Italy*

ABSTRACT

Sustainable water management has become an urgent challenge due to irregular water availability patterns and water quality issues. The effect of climate change exacerbates this phenomenon in water-scarce areas, such as the Mediterranean region, stimulating the implementation of solutions aiming to mitigate or improve environmental, social, and economic conditions. A novel solution inspired by nature, technology-oriented, explored in the past years, is constructed wetlands. Commonly applied for different types of wastewater due to its low cost and simple maintenance, they are considered a promising solution to remove pollutants while creating an improved ecosystem by increasing biodiversity around them. This research aims to assess the sustainability of two typologies of constructed wetlands in two Italian areas: Sicily, with a vertical subsurface flow constructed wetland, and Emilia Romagna, with a surface flow constructed wetland. The assessment is performed by applying a cost-benefit analysis combining primary and secondary data sources. The analysis considered the market and non-market values in both proposed scenarios to establish the feasibility of the two options and identify the most convenient one. Results show that both constructed wetlands bring more benefits (benefits-cost ratio, BCR) than costs (BCR > 0). In the case of Sicily, the BCR is lower (1) in the constructed wetland scenario, while in its absence it is almost double. If other ecosystem services are included the constructed wetland scenario reach a BCR of 4 and a ROI of 5, showing a better performance from a costing perspective than the absence one. In Emilia Romagna, the constructed wetland scenario shows a high BCR (10) and ROI (9), while the scenario in absence has obtained a negative present value indicating that the cost do not cover the benefits expected. Further research should be focused on improving ecosystem services monetary quantification from different context (i.e. rural vs urban).

## 1. Introduction

Access to water and sanitation are primary in humans' lives, as the United Nations recognises under human rights (UN, 2020). At the same time, as a scarce and stressed resource, water is crucial in producing energy and food, industry, and environmental quality. It has an enormous impact on natural resource exploitation (Sgroi et al., 2018). Moreover, the Mediterranean region faces an unstable regime and limited availability, increasing threats by climate change and drought events (WWAP, 2017).

During the past decades, multiple policies were adopted at a supranational, regional, and national scale to deal with the water-related issue. At the EU level, in 2000, the Water Framework Directive was implemented. This water policy aimed to protect water resources, ecosystems plan tailored policies to reach sustainable management of the water resources (European Commission, 2000). Other relevant policies linked to water management are the European Urban Wastewater Directive, which highlights the necessity of secondary and more severe treatment of urban wastewater in sensitive areas to protect the water resource and the environment (Djukic et al., 2016), and the EU Marine Strategy Framework Directive, adopted in 2008 and related to the improvement of marine water quality (Börger et al., 2016; EC, 1991).

\* Corresponding author at: Joint Research Centre, Ispra, Italy.
*E-mail address:* laura.garcia-herrero@ec.europa.eu (L. García-Herrero).

More recently, a new Regulation on minimum requirements for water reuse for agricultural irrigation has entered into force, with rules to be applied from June 2023 in the Circular economy action plan context, intending to stimulate and facilitate water reuse in the EU (EU, 2020).

As part of a global strategy, the European Union (EU) approved the 17 sustainable development goals, part of the 2030 Agenda for Sustainable Development. Goal number 6, "Ensure availability and sustainable management of water and sanitation for all," is related to the provision to developing countries of bilateral assistance programs and regional initiatives and in general support to the water sector, which is critical in the commitment towards to more sustainable management of water resources (United Nations, 2015).

Therefore, the role of water in the EU's policy agenda emphasizes the need to address water scarcity identifying innovative solutions to respond to raising problems.

In this framework, a promising technology for wastewater treatment that allows freshwater utilisation for alternative purposes is constructed wetland (CW). This green infrastructure mainly comprises vegetation, soil and substrates, and water, creating different mechanisms to remove contaminants or improve water quality, as natural wetlands would do (Gorgoglione and Torretta, 2018; Resende et al., 2019). Unlike grey infrastructures, CW is an easily manageable and low-cost technology that requires a minimal level of maintenance. It can be applied in different socio-economic and geographical contexts and has the flexibility to be adopted and tailored to different territorial conditions (Gkika et al., 2014; Lavrnić et al., 2018). CWs' performance is influenced by factors such as size, operating conditions and local climate, wastewaters properties, and pollution content, among others.

Due to the properties already listed and with the aim of improving the quality of water, some CW are designed to treat domestic wastewater while combining more intensive technologies to increase removal performance (Nan et al., 2020). In agriculture, for example, the inclusion this type of CW could ensure different positive effects because of its ability to block non-point sources of pollution, such as nitrogen and phosphorus, hence preventing the eutrophication phenomenon that can harm surface water bodies (Yang et al., 2020). Thus, acting as a multi-functional system that can provide several ecosystem services, such as support to the biodiversity and habitat of an environment, recreational and socio-economic services as the biomass produced could be utilised in energy production, could help in flood prevention, and control, retention of water and prevention of erosion (Lavrnić et al., 2018; Milani et al., 2019; Wang and Banzhaf, 2018).

To reveal CW's positive and negative effects in the draft and implementation of efficient policies and strategies related to water resources administration, the Water Framework Directive underlined the importance of applying appropriate economic analysis tools, such as Cost-Benefit Analysis (CBA) or Cost-Effectiveness. CBA is a wide-recognised tool to assess selected sustainability features in projects and services. It combines price flow analysis, environmental consequences by including externalities, and the social perspective of different projects or policies. Furthermore, CBA mainly adopts money or welfare as a unit of reference (Hoogmartens et al., 2014), allowing comparing the different alternative measures and scenarios - such as CW- enabling users to assess economic and financial profitability respectively a societal and a stakeholders' points of view. As recognised by Aparicio et al. (2019), the application of CBA for the evaluation of the economic feasibility of projects related to water use and reuse increased over the past few years.

This research aims to assess the sustainability of a vertical subsurface flow constructed wetland in the South of Italy (Sicily) and a surface flow constructed wetland in the North of Italy (Emilia Romagna) by applying Cost-Benefit Analysis.

## 2. Material and methods

This section presents a brief description of each case study (Table 1) and the application of the methodology, considering both market and non-market values. Further details are provided in the Supplementary Data.

**Table 1**
Main characteristics of the Catania and Bologna case studies.

|  | Sicily | Emilia Romagna |
| --- | --- | --- |
| Location | Metropolitan city of Catania | Metropolitan city of Bologna |
| Context | Urban | Rural |
| Type of green infrastructure | Retention pond + Vertical subsurface flow CW (VFCW) | Surface flow CW (SFCW) |
| Influent | Surface run-off | Agricultural drainage water |
| Flow rate | $\approx 40$ m$^3$ d$^{-1}$ (maximum 300 m$^3$ d$^{-1}$) | Varying (0–600 m$^3$ d$^{-1}$) |
| Size/Area | 1500 m$^2$ | 5557 m$^2$ |
| Scale | Full-scale | Full-scale |
| Waterproof | Yes | No |

Each case study is organized around two scenarios. First, a baseline scenario without a vertical subsurface flow CW structure is coupled to an alternative scenario with a vertical subsurface flow CW structure for Sicily. Second, for Emilia Romagna, a baseline scenario without a surface flow CW intervention is coupled with an alternative scenario with a surface flow CW intervention.

### 2.1. Sicily case study

The Sicily case study is located in the Metropolitan city of Catania, within an Ikea® land property. At the end of 2016, this area was installed a pilot-scale CW plant to treat a portion (about 2 m$^3$ d$^{-1}$) of the surface run-off collected from the retail store's parking area (Ventura et al., 2021). The University of Catania built the experimental plant within the international research joint project (Ventura et al., 2019). For the case study, the scale-up was assumed from pilot plant to full-scale CW plant in a field categorised for agricultural purposes currently used by shepherds for pasture (Fig. 1). Due to its marginalised location, a land-use change for constructive purposes is not likely.

The surface will be extended to 1500 m$^2$, composed of a retention pond occupying an area of 500 m$^2$. A vertical subsurface flow CW extended on 500 m$^2$ and an area of the relevance of about 500 m$^2$. The CW will manage the first 5 mm of rain of the parking lot of Ikea® in Catania, about 60,000 m$^2$, and will not manage unexpected events (second rainwater). It will be designed to treat a maximum flow rate of about 300 m$^3$/day (15,000 m$^3$/year). Concerning the labour force, it is not projected to have a regular staff but only periodical maintenance after its construction. However, in the case of extreme droughts, it might need specific care. The water treated will be utilised for green areas irrigation and WC flushes at the store.

Following the CBA steps, the baseline scenario in this case study is defined as the current land use without introducing the VFCW. In contrast, the alternative scenario implies applying the designed VFCW.

### 2.2. Emilia Romagna case study

This surface flow CW is located in the experimental farm of the Land Reclamation Consortium Emiliano Romagnolo Canal in Budrio (CER, according to its acronym in Italian), in Emilia Romagna. The SFCW was completed in 2000, and since 2003 has been the target of different experimental studies (Lavrnić et al., 2020b). Currently, it treats the drainage water from the experimental surroundings, a farm of 12.4 ha with different cultivation systems (mainly with trees, vegetables, and cereals). The CW has a total area of 5557 m$^2$, with a 470 m long channel and 8–10 m wide, divided into four meanders and a total capacity of 1500 m$^3$ (Lavrnić et al., 2020a). It does not have a constant inflow of water since agricultural drainage water volume mainly depends on precipitation. The main plant species are *Phragmites Australis*, *Typha Latifolia*, and *Carex* spp., Fig. 2 shows an aerial image of the case study

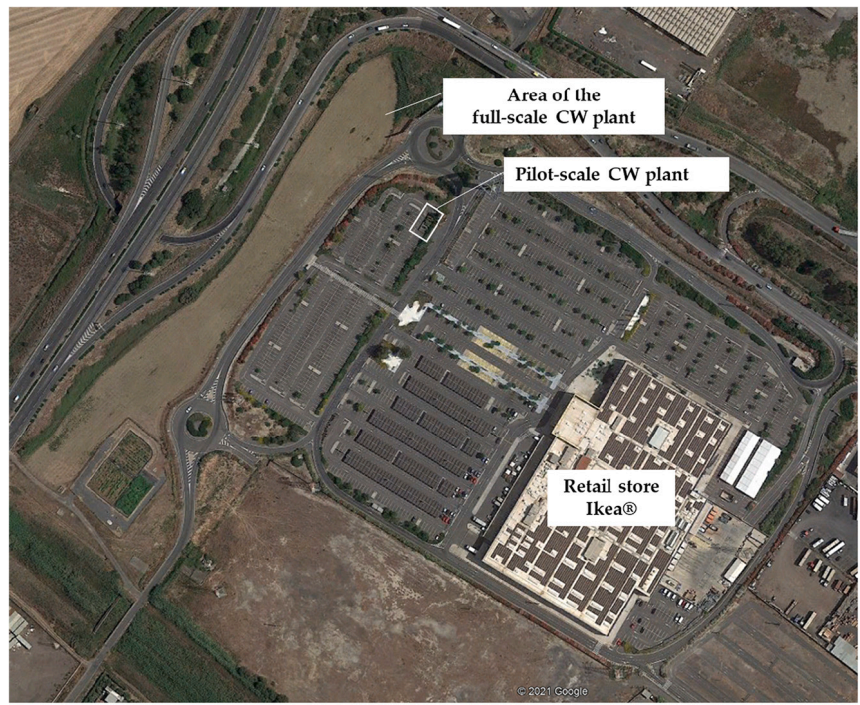

**Fig. 1.** Location of the pilot-scale CW plant and the area of the full-scale CW plant in Sicily (from Google Maps).

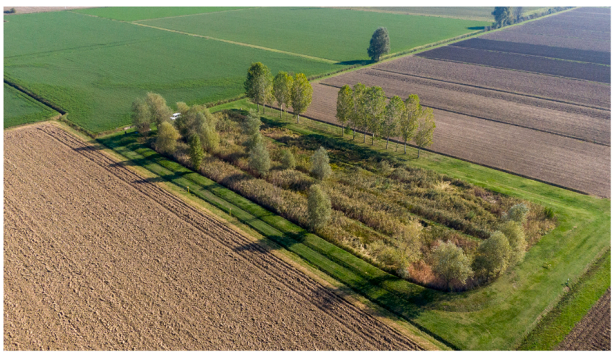

**Fig. 2.** The SFCW in Emilia Romagna and the surrounding farm area.

location.

The main ditch collects the agricultural drainage water from the fields. It flows by gravity towards the farm's end, where the SFCW is located. Two electric pumps bring the water into the wetland, as there are few meters from the main ditch to the wetland. The water flows through the different meanders, thanks to gravity and pressure from incoming water. If the water level above 0.4 m – water discharge occurs, a pump brings the water out of the SFCW to the main ditch. During the wetland construction, the labour force was concentrated on maintenance as mowing performed once a year. After almost two decades of operation, the system can still effectively treat agricultural drainage water and presents a buffer for different contaminants (Lavrnić et al., 2020b; Lavrnić et al., 2018).

The alternative scenario considered the surface currently covered with the wetland as an agricultural area with potatoes, soybeans, corn, and wheat. Therefore, these crops are selected as they are present on the remaining surface of the experimental farm.

### 2.3. Cost-benefit analysis

Cost-Benefit analysis is a widely recognised economic tool that explores the costs and the benefits of the selected project. The CBA starts from the premise that investment should only be commissioned if the benefits exceed the aggregate costs (Molinos-Senante et al., 2010), considering that the compared benefits and costs must belong to the same situation (Young and Loomis, 2014). However, as the principal limit of this methodology is identified, the fact that not all impacts can be quantified and monetised, which restricts the provided results (Huysegoms et al., 2018); while provides several advantages such as the identification of positive and negative societal cost, the inclusion of discount rate or the evidence of a general overview of impacts from different nature as evaluation (Huysegoms et al., 2018). Fig. 3 shows the steps followed to perform the CBA of the case studies based on different CBA guidelines.

Primary data collection to perform this CBA was prioritised. As both cases are under the supervision of researchers from the University of Catania, University of Bologna, and CER, each case study's responsibility was addressed in person and by phone to provide most of the information. When researchers did not have the requested information, the contact of suppliers and operational workers was facilitated. Secondary data was collected in the absence of primary data, including an extensive literature review of scientific and grey literature.

The authors' highlight as a limit of this research the utilisation of secondary data when the primary was not available and the estimations considered in some monetised items in the Emilia Romagna case since they could be outdated as the construction took place at the beginning of 2000.

Some key performance indicators (KPI) associated with CBA (step 6 in 3) were utilised to compare different dimensions in the baseline and alternative scenarios. The first step to obtain the KPIs was to set the present value or value 2019 in this research results.

PV cost = cost * $(1 + r)^{-t}$;
PV benefit = benefit * $(1 + r)^{-t}$
Equation 1. Present value.

The present value (PV), in Eq. 1, considered the discount rate (r) for future values under a specific time (t). In this research, a 5% discount rate has been applied, as widely utilised in green infrastructure assessments and recommended by the European Commission (Bixler et al., 2019; Djukic et al., 2016; European Commission, 2014; Resende et al., 2019).

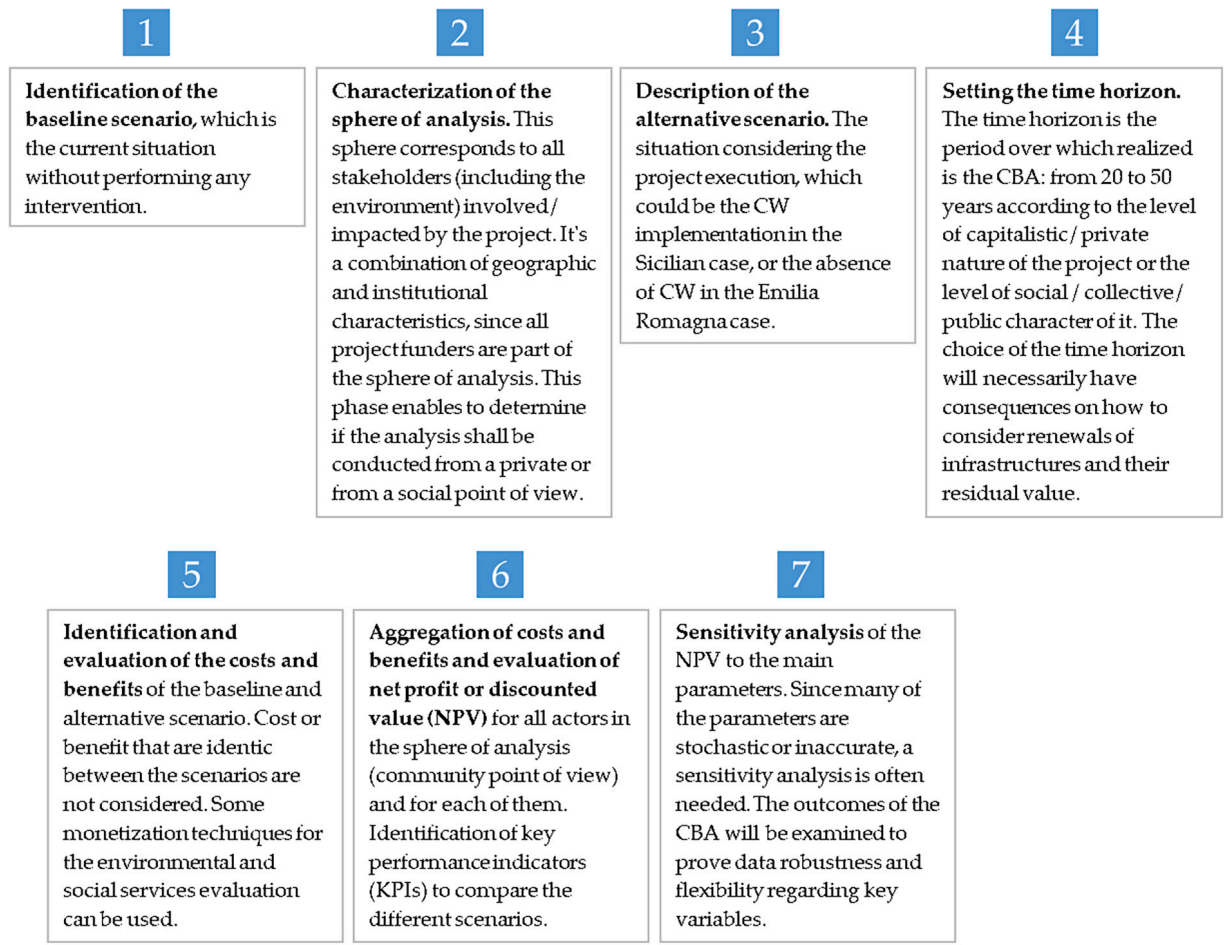

Fig. 3. Steps followed in the CBA for the two case studies.

The net present value (NPV) was essential to understand if the scenario analysed is positive or negative in terms of economic cost. At the same time, the benefit-cost ratio (BCR) offered the relationship between the relative costs and benefits of the case studies. Finally, the return on investment (ROI) is a well-known monetary indicator that measures the investment's efficiency. Eq. 2 discloses how these KPIs were obtained.

NPV = $\sum$ PV benefits - $\sum$ PV costs

Benefit Cost Ratio (BCR) = $\sum$ PV benefits / $\sum$ PV costs

Return On Investment (ROI) = ($\sum$ PV benefits – PV cost of investment) / PV cost of investment

Equation 2. NPV, BCR, and ROI formulas.

Other KPIs considered were the cost/m$^2$ and NPV/m$^2$ to test the cost referring to the size occupied by the wetland or in its absence.

## 3. Results and discussion

This section provides the results and discussion of both case studies. In addition, at the end of each case study, the results of the key performance indicators and the sensitivity analysis are provided.

### 3.1. CBA in Sicily

#### 3.1.1. Baseline scenario

Table 2 shows the results of the baseline scenario. The column "value" represents the total investment, while "value 2019" (present value) represents the current value where a discount rate of 5% (when it applied) and present in Table 3. This discount rate has been selected as is the one recommended by the European Commission for the cost-benefit analysis and widely applied in other studies analysing wetlands

Table 2
Cost from the baseline scenario in Sicily without VFCW.

| Costs | Value | Unit | Source |
|---|---|---|---|
| CAPEX (capital costs) | | | |
| Purchasing land cost – 1500m$^2$ | 5000 | € in Year 0 | Interview |
| OPEX (operational and maintenance) | | | |
| Financial | | | |
| Land insurance | Not expected | €/year | Interview |
| Plant insurance | Not expected | €/year | Interview |
| Maintenance | | | |
| Land maintenance (weeds removal labour) | Not expected | €/year | Interview |
| External costs | | | |
| Environmental costs (flood risk) | 2369 | €/year | (i) |
| Alternative water treatment methods or water sources costs | | | |
| Outsourcing costs for water treatment | Not expected | €/year | Interview |
| Grey infrastructure plant cost | Not expected | €/year | Interview |
| Energy costs for grey infrastructure | Not expected | €/year | Interview |
| Disposal cost of greywater | Not expected | €/year | Interview |
| Taxes | | | |
| Sewer tax for greywater | Not expected | €/year | Interview |

i. USD$0.18/m$^3$ in 2018 (Nordman et al., 2018). It was updated with inflation and $/€ to reflect the value in 2019. It considers the maximum flow rate that the GI can treat (15,000 m$^3$).

Table 3
Benefits from the baseline scenario in Sicily without VFCW.

| Benefits | Unit | Value | Value 2019 | Source |
| --- | --- | --- | --- | --- |
| Alternative use of the land | / | | | Interview |
| No material and equipment cost for GI | € | 90,500 | 3016.67 | Green Infrastructure |
| No labour cost for green infrastructure | € | 12,500 | 417.67 | Green Infrastructure |
| No maintenance cost for green infrastructure | €/year | 5000 | 5000 | Green Infrastructure |
| No electrical energy for green infrastructure | €/year | 940.8 | 940.80 | Green Infrastructure |
| No mowing and disposal of vegetable biomass cost | €/year | 1000 | 952.38 | Alternative scenario |
| Non-maintenance benefit because of pasture use of the land (Estimated 3 interventions/year 2 h/intervention) | €/year | 540 | 540 | (i) |

(i) 90€/h for 1500 m2 (UNCAI, 2019).

(Molinos-Senante et al., 2010; Hernández-Sancho and Sala-Garrido, 2009; European Commission, 2014). The lifespan of the green structure is 30 years, which is the estimated lifetime of the constructed wetland (Alves et al., 2019).

In the first instance, the baseline condition has only one initial investment: the land purchase cost. The operational costs are represented by the maintenance, which is zero, given that the land is freely granted for pasture. The environmental costs are composed only by the flood risk, considering the absence of action. No other water treatment cost was identified in this section, as no taxes are paid to manage sewer greywater.

The largest burden is allocated to external cost due to environmental flood management, representing the total cost of the baseline scenario.

Table 3 shows the different expected benefits of the baseline scenario. Those sources under the Green Infrastructure label are disclosed in the alternative scenario of the VFCW.

According to the calculated benefits figures, about 10,900 €/year are expected to be obtained when no action is taken. The largest contributor is those benefit items under the green infrastructure construction (initial cost and maintenance).

*3.1.2. Alternative scenario: Constructed wetland*

Table 4 shows the cost calculated to implement the VFCW from its initial investment to its maintenance.

Under the alternative scenario, a larger initial investment cost is evident compared to the baseline, as the land must be prepared to allocate a VFCW. The primary cost item refers to the core works of the VFCW. The total cost per year is approximately 10,500 €/year. About 34% is associated with the initial investment cost, and about 66% is related to operational costs. The CW in Catania does not have a financial cost associated with land-use typology. Its utilisation does not require any type of insurance. There are no-cost outcomes obtained from nutrient and water distribution. On the one hand, there is a modest amount of nutrients present. On the other hand, currently, the water treated cannot be capitalised profitably.

When focusing on the environmental cost, the construction stage represents almost 100% of the overall environmental impacts due to excavations. This value is larger than the 80% specified in other studies (Resende et al., 2019). This cost has been calculated by establishing a carbon price that is highly volatile under current environmental policies and the carbon market. The large environmental burden happens once during the VFCW construction stage. Therefore, it could be neglected as these structures' lifespan is extended. In Table 5, the expected benefits of the alternative scenario are disclosed.

A large benefit has been associated with the water output, expecting to be included in the water system and profit from it. Considering the

Table 4
Cost expected from the alternative scenario in Sicily, with VFCW.

| Expected costs | Unit | Value | Value 2019 | **Source** |
| --- | --- | --- | --- | --- |
| **Initial Investment** | | | | |
| Purchasing land cost – 1500m$^2$ | € | 5000 | 166.67 | Interview |
| *Materials and Equipment* | | | | |
| Plant cost | € | 5700 | 190 | Interview |
| Excavation cost | € | 9649 | 321.67 | Interview |
| Ponds waterproofing | € | 34,000 | 1133.33 | Interview |
| Electrical system | € | 9000 | 300 | Interview |
| Hydraulic system | € | 9000 | 300 | Interview |
| Completion work of ponds (non-woven fabric, bio jute net, filling ground, inert material) | € | 21,250 | 708.33 | Interview |
| Lifting pumps | € | 1900 | 63.33 | Interview |
| *Labour* | | | | |
| Completion work of ponds labour | € | 3750 | 125 | Interview |
| Ponds waterproofing labour | € | 6000 | 200 | Interview |
| Electrical system labour | € | 1000 | 33.33 | Interview |
| Hydraulic system labour | € | 1000 | 33.33 | Interview |
| Planting labour | € | 300 | 10 | Interview |
| Lifting pumps labour | € | 100 | 3.33 | Interview |
| Excavation labour | € | 350 | 11.67 | Interview |
| **Operational Costs** | | | | |
| *Fixed* | | | | |
| Electrical energy 400 kwh/month | €/year | 940.80 | 940.80 | (i) |
| Fixed plant staff cost | €/month | It is not needed | | Interview |
| Plant monitoring cost | €/month | It is not monitored | | Interview |
| *Variable* | | | | |
| Reagent substances (sludge) | €/year | No sludge - modest concentrations (<1 mg/l) | | Interview |
| Mowing and disposal of vegetable biomass cost | €/year | 1000 | 952.38 | (ii) |
| Ordinary maintenance of the plant | €/year | 5000 | 5000 | Interview |
| Irrigations (when necessary) | €/year | It is not needed | | Interview |
| **Financial costs** | | | | |
| Land insurance | €/year | There is not land insurance | | Interview |
| Plant insurance | €/year | There is not plant insurance | | Interview |
| **Taxes** | | | | |
| Plant taxes | €/year | There is not plant taxes | | Interview |
| Water distribution (if sold) | €/year | There is not water sold | | Interview |
| Nutrients distribution (if sold) | €/year | There are not nutrients sold | | Interview |
| **External costs** | | | | |
| Social costs | €/year | It is not monitored. It could be reputational for the company. | | |
| Exceptional irrigations | €/year | This cost is not expected | | |
| Environmental costs | €/m3 | 43.42 | 1.45 | (iii) |

i. 0.196 €/kw provided by Ikea®.
ii. Interview, vegetable biomass mowing, and disposal is a cost to sustain after one year. Therefore, it was discounted at the 5% rate. This cost includes the maintenance of the use of the land.
iii. Environmental costs are represented only by the $CO_2$ of excavations, which were valued using the kg of $CO_2$ equivalent/m$^3$ of the excavation($\simeq$0.48) (Wernet et al., 2016), the m$^3$ of the land (1500 m$^3$) and the cost of $CO_2$ (60 €/ton) (Eff. Carbon Rates 2021, 2021).

location of the wetland, in an island suffering from high temperatures and irregular raining patterns, the cost of the water reused has been established in 0.9€/m3, while this cost in Italy could rank from 0.25 to 1.5€/m3 (Pistocchi et al., 2018). Improve air quality is referred to as reduced air pollution, thanks to the $CO_2$ uptake by these natural ecosystems. This figure might be more prominent if other externalities such as human health improvement due to pollution-related diseases are considered. Several studies highlight the role of ecosystem services and their monetary value, a relevant study from Costanza et al. (2014) quantified that the land use changes at global level between 1997 and

**Table 5**
Benefits expected from the alternative scenario in Sicily, with VFCW.

| Expected benefits | Unit | Value | Value 2019 | Source |
|---|---|---|---|---|
| Output water 12,000 m³/year | €/m2 | 10,800 | 10,285.71 | (i) |
| Improve air quality and $CO_2$ storage | €/year | 15,884.12 | | (ii) |
| Reduce the risk of flood | €/year | 2369 | | (iii) |

i. 0.9€/m3 water reuse cost in Italy (Pistocchi et al., 2018).
ii. 1.48USD$/m³ in 2018 (Nordman et al., 2018). It has been updated with inflation adapted to the European currency in 2019. Air pollution reduction + $CO_2$ storage.
iii. 0.18 USD$/m³ in 2018 (Nordman et al., 2018). It has been updated with inflation adapted to the European currency in 2019.

2011 ensued in a loss of ecosystem services of between $4.3 and $20.2 trillion/yr. It is also highlighted the difficulty for general public to understand the value of the services this ecosystem could bring, such as reduction of loss of resources, protection of human health, nutrients recycling and restoration and reuse of water resources, if the land is expected to be used for further proposes (Masi et al., 2018; Resende et al., 2019). Aesthetical benefits, such as improve aesthetic value, scenic beauty, temperature refreshment, have not been considered as the area is not easily accessible. Still, it should be considered in the future together as depending on the context it could offer positive welfare effects (benefits) (Darnthamrongkul and Mozingo, 2021; Jensen et al., 2019; Ureta et al., 2021). Moreover, the value of wetlands per biome in monetary units among different ecosystem evaluated has been revelled the highest – mainly under mangroves (de Groot et al., 2012).

### 3.1.3. Performance and sensitivity analysis

Key Performance Indicators (KPI) considered to evaluate both scenarios are disclosed in Table 6.

Results evidence that the NPV is higher in the baseline scenario due to the present net value obtained, while at long term the alternative scenario could become more profitable. Following the CBA premise that benefits should extend the costs, constructed wetlands could be suitable for this case study, but the ROI and BCR is low. If considering an increase in the cost of the reuse water (from 0.9€/m3 to 1€/m3), and aesthetical value which is feasible due to its socio-economic and geographical context (>50 km from urban areas and the number of visitors to this parking lot, the BRC could reach almost 5 (adjusted with inflation 2.98 $/m3 from Nordman et al., 2018), and the ROI almost 3, being superior than in the baseline scenario. Therefore, further research should be made to better capture the costing service of this ecosystem.

For both cases, a sensitivity analysis was carried out to test how robust the results are by modifying selected inputs. In this case, the selected variables to test due to their influence on the results are:

- The discount rate. The reference value was modified by 2.5 and 7.5%, as the discount rate is often tested in similar studies (Alves et al., 2019; Molinos-Senante et al., 2010).

**Table 6**
KPI comparing the alternative and baseline scenario in the case study in Sicily.

| KPI | VALUE |
|---|---|
| *Alternative scenario: Green Infrastructure* | |
| Net Present Value (NPV) | 2.46 |
| Benefit-Cost Ratio (BCR) | 1.00 |
| Return On Investment (ROI) | 0 |
| Total Costs/m² (1500m²) | 7.00 |
| NPV/m² (1500 m²) | 0 |
| | |
| *Baseline scenario* | |
| Net Present Value (NPV) | 5838.59 |
| Benefit-Cost Ratio (BCR) | 2.16 |

- The ordinary maintenance of the plant, which represents a high yearly cost, can vary based on the necessity, modifying the reference value by 3000 and 7000.
- The output water in m3/year varies on precipitation base, modifying the reference value by 6000 and 15,000 m3/year (as the structure has been projected to deal with a maximum flow rate of 15,000 m3/year).
- The sensitivity analysis results can be observed by changing the discount rate values, ordinary maintenance of the plant, and the output water in m3/year; there is only a minimum variation in the totals and the KPIs.
- The NPV is forecasted over 30 years (from 2019 to 2049) with an interval confidence bound of 95% based on the minimum and maximum possible variation of output water. The NPV is forecasted to be negative only in case output water values are near the minimum flow rate.

### 3.2. CBA in Emilia Romagna

#### 3.2.1. Baseline scenario

In this case study, the baseline scenario is the SFCW. In contrast, the alternative scenario was defined as a theoretical cultivation rotation of potatoes, soybeans, corn, and wheat. Table 7 shows the cost from the baseline scenario, considering that the area has 0.55 ha and can treat up to about 20,000 m3 of water/year depending on the rate of annual precipitations In this study, an overall inflow and outflow of 16,186 m³/year and 7119 m³/year were considered as a mean value of volume of treated wastewater in the years 2018 and 2019 (Lavrnić et al., 2018).

Table 7 shows that land purchase has not been included in the cost as the CER owns this land for >20 years. Therefore, it was deemed appropriate to calculate the opportunity cost deriving from the income related to the following crops-, potato, soybeans, corn, wheat, which are usually grown in a rotation system. The excavation costs were calculated for a depth of 0.40 m from the field level and 0.90 m including embankments and the inclusion of labour costs for the equivalent of 2 working days; the cost of the electrical system includes the costs of bringing electricity from the rural buildings present at the entrance to the farm, up to the Phyto-depuration area. Since the area is subject to scientific research activities, the macrophytes planted were taken from natural environments and introduced into the area by carrying out several tests to test their engraftment. To calculate their cost, reference is made to the total sale price of the seedlings in multipots of 60 units, planted with a crop density equal to 1 unit / m²; no base or cover fertilisations were performed, as the plant essences were selected for their high rustic characteristics; the hydraulic system costs have been calculated based on the prices provided by the interview performed to Impianti Bragaglia at the end of 2019. The total value of the investment costs expressed reach about per year in 30 years of life.

The operational costs are mainly fixed—the first item related to maintenance, which is 1720€ and takes place once every five years. The ordinary maintenance is expected to happen 5 times a year, each time with a duration of 2 working days. The energy costs are related to the electric pump functioning to transfer the water from the main channel to the constructed wetland, while no pumping is needed to transfer clean water from the constructed wetland to the main channel once the water has been phyto-depurated as it works thanks to gravity. There is a cost item related to labour – 2 working days per year - when mowing and harvesting the constructed wetland biomass. The IMU cost, an Italian Municipal Property Tax related to property land, was estimated as 3.5% on the total taxable amount, corresponding to the surface of the phyto-depuration area on the farm's total area. The environmental cost was always related to excavation costs. Overall costs reach 2121.28€/year.

This cost could be reduced if better performance excavators were utilised for the wetland construction consuming fewer fossil fuels. Table 8 indicates the benefits of the current scenario.

Expected benefits have been calculated by applying estimations

Table 7
Cost from the baseline scenario in Emilia Romagna, with SFCW.

| Costs | Unit | Value | Value 2019 | Source |
|---|---|---|---|---|
| **Initial investment** | | | | |
| Opportunity cost land | €/ha per year | | 25.20 | Alternative Scenario |
| Design cost | € | 2000 | 66.67 | Interview |
| Cost of excavation and embankment (including labour) | € | 3500 | 116.67 | Interview |
| Electrical system cost | € | 500 | 16.67 | Interview |
| Cost of basic fertilization | €/kg | 0 | 0 | Interview |
| Cost of seedlings (in 60 holes multipot pot) | € | 2800 | 93.33 | Interview |
| Irrigation cost | € | 3708 | 123.60 | (i) |
| Other costs (concrete structures) | € | 5000 | 233.33 | Interview |
| Labour cost (assembly) | € | 560 | 18.67 | Interview |
| Cost of submersible electric pumps (2) AP.50.11.3 1KW, 380 V, 3A | € | 2000 | 66.67 | Interview |
| Cost of pipes 5€ / m² | € | 500 | 16.67 | Interview |
| Cost of electric box | € | 300 | 10 | Interview |
| Volumetric impulse meter (2) | € | 660 | 22 | Interview |
| Level sensor (2) | € | 1700 | 56.67 | Interview |
| **Operational cost** | | | | |
| *Fixed* | | | | |
| Maintenance costs (extraordinary) | €/year | 344 | 344 | Interview |
| Ordinary plant maintenance (green management) | €/hour | 18 | 1080 | Interview |
| Energy costs 138 kW/month (ordinary operation) | €/year | 46.32 | 46,32 | Interview (ii) |
| *Variable* | | | | |
| Labour for mowing and dry biomass harvesting | €/hour | 18.00 | 288 | Interview |
| Irrigation | | | | |
| Planting of new vegetation/plants | | | | |
| **Financial cost and taxes** | | | | |
| Plant insurance | Not applied | | | |
| IMU (tax) | €/ha | 200 | 70 | Interview (v) |
| **External cost** | | | | |
| Social cost | | | | |
| Environmental cost | € | 44.80 | 1.49 | (vi) |

i. 1.20€/m² for 3090 m³ Gruppo Hera, 2019 (non-domestic use, agricultural purposes).
ii. Retrived from CER electrical bill (0.0125 €/kW + fixed costs = 0.20€/kW). Supplied by Nova AEG.
iii. 3.5% of the total taxable amount corresponds to the surface of the phyto-depuration area on the farm's total area.
iv. It refers to the $CO_2$ emissions produced by excavation ($\simeq$0.48t$CO_2$eq.) multiplied by the market price of $CO_2$eq./ton emissions (60 € / ton) (Eff. Carbon Rates 2021, 2021) already established in different studies located outside Italy as other studies have not been found. This research gap suggests, as in the case of Sicily, the need to better explore the role of this ecosystem service in this case, in a rural area.

### 3.2.2. Alternative scenario: Without wetland

There is no opportunity cost in the alternative scenario as the crop field is used for cropping purposes. The following rotation crops have been considering: potato, wheat, maize and soybeans. Operating costs are related to standard agronomic cultivation practices' average costs and are shown in Annex 1. The value of the IMU remains unchanged concerning the condition in which the CW is present. The alternative scenario costs amount to 2266€ without including external cost, while when it is included, it reaches 2296€. Table 9 discloses the expected cost from the alternative scenario.

Yearly benefits of the alternative scenario reach over as is evidenced in Table 10 due to the economic benefit of selling the crops and

Table 8
Benefits from the baseline scenario in Emilia Romagna with SFCW.

| Benefits | Unit | Value | Value 2019 | Source |
|---|---|---|---|---|
| Lower P pollution in water | €/year | 10.14 | 10.14 | (i) |
| Lower N pollution in water | €/year | 123.54 | 123.54 | (ii) |
| TSS reduction | €/year | 298.90 | 298.90 | (iii) |
| Ecosystem benefits | €/year | 17,599.07 | 17,599.10 | (iv) |
| Flood reduction | €/year | 30.10 | 30.10 | (v) |
| Agricultural benefits | €/year | 2554.01 | 2432.39 | (vi) |
| Scenic amenity value | €/year | 654.93 | 654.93 | (vii) |

i. 109 kg/year removal from Lavrnić et al., 2018 considering 29250AUS$/t in 2012 (Daniels et al., 2012). The figure has been updated with inflation adapted to the European currency in 2019.
ii. 0.36 kg/year removal from Lavrnić et al., 2018 considering 861 AUS$/t in 2012 (Daniels et al., 2012). The figure has been updated with inflation adapted to the European currency in 2019 (Daniels et al., 2012).
iii. 1.36 USD$/m³ (Nordman et al., 2018). It has been updated with inflation adapted to the European currency in 2019
iv. 24,056 AUS$/ha considering the minimum value (Daniels et al., 2012). The figure has been updated with inflation adapted to the European currency in 2019.
v. 0.18USD$/m³ (Nordman et al., 2018). It has been updated with inflation adapted to the European currency in 2019.
vi. Agricultural benefits from the potential reuse of wastewater considering an 'extra' income for avoided losses on production due to drought events equal to 20% of Gross Saleable Production (4036.05 €) of a field of $\simeq$ 3,00 ha (Verlicchi et al., 2012).
vii. 2.98USD$/m³ (Nordman et al., 2018). It has been updated with inflation adapted to European currency in 2019.

Table 9
Cost expected from the alternative scenario in Emilia Romagna, without SFCW.

| Expected costs | Unit | Value | Value 2019 | Source |
|---|---|---|---|---|
| **Initial Investment** | | | | |
| Opportunity cost land | €/year | It is no expected | | Interview |
| **Operational cost** | | | | |
| Potato/wheat/corn/soybean cultivation | €/year | 2196.12 | 2196.12 | (i) |
| **Financial cost and taxes** | | | | |
| IMU (tax) | €/ha | 200 | 70 | Interview (ii) |
| **External cost** | | | | |
| Risk of flood | € | 30.10 | 30.10 | (iii) |

i. Centro Ricerche Produzioni Vegetali (CRPV) and inflated to 2019 (World Bank).
ii. 3.5% on the total taxable amount, corresponding to the surface of the Phyto-depuration area (in this scenario, crop) on the farm's total area.
iii. 0.18USD$/m³ (Nordman et al., 2018). It has been updated with inflation adapted to European currency in 2019.

Table 10
Benefit expected from the alternative scenario in Emilia Romagna, without SFCW.

| Expected benefits | Unit | Value | Value 2019 | Source |
|---|---|---|---|---|
| Gross saleable production | € | | 2219.83 | (i) |
| No $CO_2$ emission due to excavation | € | 44.80 | 1.49 | (ii) |

i. Retrieved from ISTAT (yield q/ha) and ISMEA (price €/t) and inflated to 2019 (World Bank).
ii. It refers to the $CO_2$ emissions produced by excavation ($\simeq$0.48t$CO_2$eq.) multiplied by the market price of $CO_2$eq./ton emissions (60€/ton), (Eff. Carbon Rates 2021, 2021).

environmental benefit (externality) since no excavation like the one in the baseline scenario would be required. Some authors (Baldocchi, 2003; Bondeau et al., 2007) associate carbon storage – sequestration - to specific crop production due to carbon intake naturally occurring in certain agro-systems. This item has not been included in this research, as it is not uniformly recognised in the scientific community. It could imply a zero balance once the crop is harvested.

### 3.2.3. Performance and sensitivity analysis

The analysis of KPIs in both scenarios in Budrio, Emilia Romagna is disclosed in Table 11.

Results show that the NPV is higher in the baseline scenario than the SFCW due to the high benefit-cost ratio obtained. Following the CBA premise, that benefit should extend the costs, therefore constructed wetlands could be a suitable structure to consider in this case study. The ROI value is also encouraging in the SFCW scenario and shows a fast repayment of the costs.

The BCR shows that the baseline scenario's benefits are about higher than the costs, as the value is >0. In comparison, the alternative scenario brings more than benefits higher than the costs, around 9 times.

In this case study, also a sensitivity analysis was carried out on the system by varying the discount rate as it was done in the case of Sicily.

The sensitivity analysis results can also be observed in a way that changed the discount rate values, the output water $m^3$/year drastically, and the value of the biomass produced. There is only a minimum variation in the totals and the KPIs.

The NPV is forecasted over 30 years (from 2019 to 2049) with an interval confidence bound of 95% based on the minimum and maximum possible variation of output water. The NPV is forecasted to remain positive when the output water values are near the minimum flow rate.

### 3.3. Case studies comparison

In both case studies, the presence of wetlands has more benefits than cost. In the case of Sicily, the scenario without the wetland brings more benefits due to the cost involved in the construction of the wetland. Instead, if additional benefits such as aesthetical values are included in the scenario with the wetland, it could become a promising scenario reaching a very high ROI and BCR (higher than the scenario without VFCW). In the case of Emilia Romagna, the wetland scenario shows a better costing performance than the alternative scenario. The benefits from selling the crop if the surface currently occupied by the wetland are not worthy when comparing all benefits can be obtained from the wetland scenario. In fact, the NPV in the alternative scenario is negative, and the BCR is <1, therefore it is not recommended to perform that scenario. A key aspect should be further explored in further research is the need to differentiate benefits humans can attribute to this system from rural and urban areas, while also other benefits associated to the biodiversity improvement (or loss avoidance) could be also explored. When reviewing the methodology applied, a lack of CBA is conducted on this typology nature-based solutions, which is difficult to compare with other studies. Therefore, as recommended by the European Commission (European Commission, 2014), this methodology should be widely utilised to support decision making to move towards the decarbonisation plan expected in The Green Deal (European Commission, 2019) while being aligned with different SDGs, beyond the SDG 6. Table 12. Total cost, benefits and NPV in selected cases studies. Shows the total cost, benefits and NPV from each case study.

### 4. Conclusions

The current study presents a CBA of two types of constructed wetland in two Italian locations: a vertical subsurface flow constructed wetland in the south of Italy, in Sicily, and a surface flow constructed wetland in the north of Italy in Emilia Romagna. The CBA methodology allowed to compare a constructed wetland scenario with a scenario without this intervention, offering numerical cost and benefits of each option. The outcomes of this research evidence that both types of constructed wetlands represent promising results in terms of their cost. In Sicily, the BRC is positive but low (ratio equal to 1), showing that it brings more benefits than costs. While comparing with the absence of CW scenario still this last one has more benefits due to the lack of investments (NPV around 6000€/yr and a BCR around 2). If further ecosystem services (mainly aesthetical are included), the wetland scenario could reach a BRC of 5 with a ROI of 4 (considering secondary data from studies outside Italy), with further incomes entering for example with a ticket to visit the structure or to enjoy the area. In that case, the wetland scenario should be prioritised from a costing perspective because 1) the investment will be repaid very fast as the very high ROI (8.96) indicates, 2) the NPV is higher (around 19,000€/year) compared with the negative value in the alternative scenario (−75€/year), where the negative value indicates that there are more cost than profits, therefore it is not recommended to change the current status.

**Table 11**
KPI comparing the alternative and baseline scenario in the case study in Emilia Romagna.

| KPI | VALUE |
| --- | --- |
| Baseline scenario: Green infrastructure | |
| Net Present Value (NPV) | 19,027.79 |
| Benefit-Cost Ratio (BCR) | 9.97 |
| Return On Investment (ROI) | 8.97 |
| Total Costs/$m^2$ | 0.38 |
| NPV/$m^2$ | 3.424 |
| | |
| Alternative scenario | |
| Net Present Value (NPV) | −74.9 |
| Benefit-Cost Ratio (BCR) | 0.97 |

**Table 12**
Total cost, benefits and NPV in selected cases studies.

| | Sicily | | Emilia Romagna | |
| --- | --- | --- | --- | --- |
| | Without CW | With CW | With CW | Without CW |
| Total Costs | 5027.89 | 10,494.59 | 19,290.80 | 2426.22 |
| Total Benefits | 110,479.80 | 11,011.34 | 21,270.69 | 2264.63 |
| NPV | 5838.59 | 2.46 | 19,027.79 | −74.90 |

Further research could be driven to explore other social (human preferences) and environmental benefits (such as biodiversity) of these structures. Additionally, other sustainability assessment techniques, such as those under the life cycle thinking method, could be applied to bring a systemic approach. Constructed wetland could bring new business model development under a favourable policy framework linked with current trends about the circular economy. Thus, an exploratory analysis of business model design, including this infrastructure, could be relevant for moving towards sustainability.

### Declaration of Competing Interest

The authors declare that they have no known competing financial interests or personal relationships that could have appeared to influence the work reported in this paper.

### Acknowledgements

This research has been performed under GREEN4WATER, a national research project coordinated by the University of Bologna and funded by the Ministry of Education, University and Research within the PRIN 2015 call (grant number: PRIN2015AKR4HX) available at https://site.unibo.it/green4water.

This work was also supported by the WATERAGRI Project (water retention and nutrient recycling in soils and streams for improved agricultural production), which received funding from the European Union's Horizon 2020 research and innovation programme under grant agreement No 858375.

Special thanks to Giusi Crisafulli and Ionut Alexandru Vartolomei, who supported data collection; as well as Mr. Stefano Anconelli and the Canale Emilia Romagnolo team for their precious time facilitating data and solving doubts.

**Appendix A. Supplementary data**

Supplementary data to this article can be found online at https://doi.org/10.1016/j.ecoleng.2022.106797.